*Article*

# Ideal Node Enquiry Search Algorithm (INESH) in MANETS

**Mohammad Riyaz Belgaum[1,*], Safeeullah Soomro[1], Zainab Alansari[1,2] and Muhammad Alam[3,4]**

[1]College of Computer Science, AMA International University, Kingdom of Bahrain
(bmdriyaz‖s.soomro‖zeinab)@amaiu.edu.bh
[2]University of Malaya, Kuala Lumpur, Malaysia.
z.alansari@siswa.um.edu.my
[3]College of Computer Science and IS (IOBM), Pakistan
malam@iobm.edu.pk
[4]Institute of Post Graduate Studies, Universiti Kuala Lumpur (UniKL IPS), Malaysia
*Correspondence: bmdriyaz@amaiu.edu.bh

**Abstract:** The different routing protocols in Mobile Ad hoc Networks take after various methodologies to send the data starting from one node then onto the next. The nodes in the system are non-static and they move arbitrarily and are inclined to interface disappointment which makes dependably to discover new routes to the destination. During the forwarding of packets to the destination, various intermediate nodes take part in routing, where such node should be an ideal node. An algorithm is proposed here to know the ideal node after studying the features of the reactive routing protocols. The malicious node can be eliminated from the networking function and the overhead on the protocol can be reduced. The node chooses the neighbor which can be found in less number of bounces and with less time delay and keeping up the QoS.

## 1. Introduction

An arrangement of versatile nodes that perform essential systems administration works in an infrastructure less environment is said be a portable specially appointed system (Figure1: MANET). Nodes that fall in the correspondence run speak with each other and which don't come in the range take after the idea of multi-hop[1] for correspondence. In the system every node assumes a double part as a host by the sending and as a switch in directing packets to the destination.

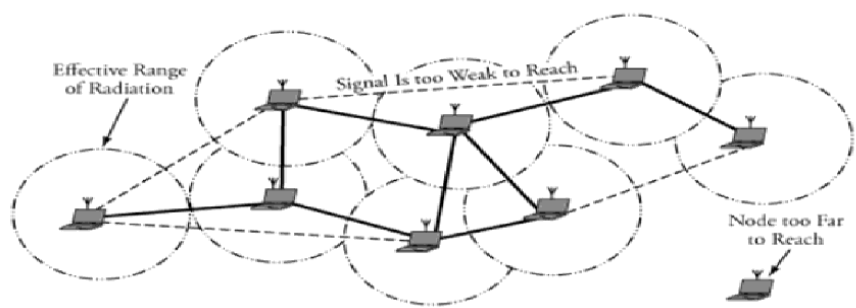

**Figure 1.** MANET.

Keeping up security is an essential capacity of any of the routing protocol in each period of the network function[1]. In light of the non-static topological conduct of the system and due to being the system open which enables the system to develop and contract because of expansion and erasure of the nodes whenever gives chance for the interloper nodes to bother the typical directing procedure. What's more, if there does not exist a typical guaranteeing authority for validating and ensuring the nodes then a dependable transmission is unrealistic.

*1.1 Security Related Goals and Challenges*

Security administrations are expected to ensure that the information is exchanged over the system with unwavering quality and furthermore keeping the assets of the framework ensured. To accomplish the targets, the arrangements of security administrations are: availability, confidentiality, authenticity, integrity and non-repudiation [2, 3].

1.1.1. Availability

Though the framework is experiencing different issues like with data transmission, network however the accessibility benefit guarantees that still the assets are accessible in an auspicious way. The unsafe impacts of accessibility of a system are asset consumption assaults and bundles dropping proportion.

1.1.2. Confidentiality

The data winning in the system is not to be shared among every single unapproved node and this is accomplished by Confidentiality. So as to accomplish Confidentiality numerous encryption methods can be utilized to influence just to the approved nodes can share the transmission of data and the private and public keys.

1.1.3. Authenticity

To demonstrate a node as an honest to goodness client the system benefit utilized is Authenticity. The nonattendance of this administration can influence any node in the system to imitate any node, and afterward having an aggregate control catch and control over the total system.

1.1.4. Integrity

The information which is been transmitted in the system can be altered either intentionally or in some cases unintentionally. The Integrity arrange benefit guarantees that the data which is been transmitted is not changed.

1.1.5. Non-Repudiation

This administration ensures that the message transmission hosts been done between the two gatherings and it can't be denied. Additionally utilizing this administration it helps in recognizing and confining of traded off nodes in the system.

Imparting through the system in sheltered and secure way has been a testing errand in light of
- Not being a steady foundation.
- The interfaces in the system are inclined to break and not secure.
- Scarcity or over-burden on the framework resources
- The arrange topology being dynamic.

In this examination the arrangement of the exploration is to consider the different reactive routing protocols in MANETS and break down the perfect node to forward the packets in the course to achieve goal. The proposed approach will be utilized to upgrade the current reactive routing protocol by taking out the malicious nodes by utilizing any of the officially characterized system to dispense with them and locate the perfect node from the accessible nodes to enhance security in while the system capacities are completed. This exploration will principally concentrate on different reactive routing protocols and the threats and attacks on them. The purposes behind security threats are considered for giving an answer to address the difficulties of security in the system and do standard system operations in a secured way.

1. What are the diverse security threats for the reactive routing protocols?
2. What are the explanations behind the threats?
3. Strategies to make the system solid and secured.

The researcher focuses on the reactive routing protocols. The attacks and threats on these routing protocols are examined, because of security such malicious node will be deleted from participating in sending the packet. Presently the explanations behind these sorts of attacks and threats are examined and the issues in this technique are considered for look into. This examination will be an enormous commitment in the range of MANETS to enhance security by using the proposed Ideal Node Enquiry Search (INESH) algorithm in the reactive routing protocols while sending the packets from source to destination. This could fill in as reference for different scientists to upgrade other class of routing protocols in view of their conduct to enhance security in MANETS.

*1.2 Conceptual Framework / Theoretical Framework*

The investigation of working procedure of each of the reactive routing protocol alongside the attacks and threats on them is examined. There are different directing systems embraced by various conventions when there are assaults on the conventions. An investigation of purposes behind the attacks and threats will be directed which makes to introduce the INESH algorithm to upgrade security in the revelation of courses and upkeep of courses in this exploration.

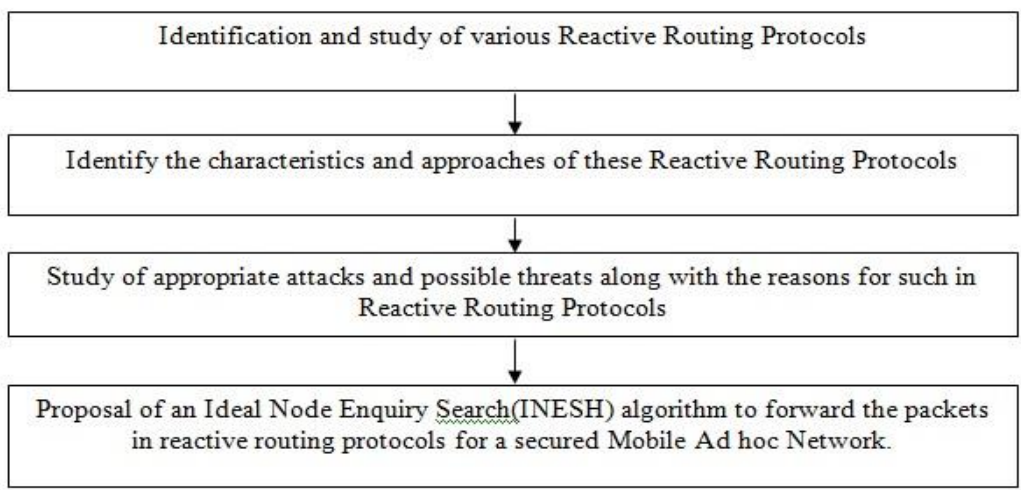

**Figure 2.** Conceptual Framework.

## 2. Literature Review

There are many sorts of routing protocols that have been intended for MANETS [4-8]. Furthermore, the features of the routing protocols are examined from [9].These protocols can be classified as follows.

| Routing Protocols | | | | |
|---|---|---|---|---|
| Topology based | | | Position based | |
| Proactive | Reactive | Hybrid | Greedy Forwarding | Restricted Flooding |
| DSDV OLSR TBRPF | DSR AODV TORA | ZRP CGSR HSR | SELAR NFP MFR | RTCP ARP LAR DREAM |

**Figure 3.** Routing Protocols.

*2.1 Reactive Routing Protocols*

2.1.1. Dynamic Source Routing Protocol

The Dynamic Source Routing protocol is a reactive routing protocol which communicates following two phases namely route discovery and route maintenance [10]. Initially the routes are discovered for transmitting the packets between source node and the target node. A route cache is maintained having the information of the recently used routes. Because of the dynamic changes in the topology, there is a chance of routes being broken in the route maintenance phase still it ensures that the packet is safely transmitted to the target. More over the researcher simulated the protocol using NS-2 to focus on the performance of this protocol using the metrics like packet delivery ratio. It was found that the packet delivery ratio is more when there were malicious nodes than in absence of malicious nodes.

2.1.2. Ad hoc On Demand Distance Vector Protocol

The AODV protocol explains its functionality [11]. It is stated that the features of both DSR and DSDV are combined. The author explains the working of AODV protocol along with the Black hole attack and its effect on the protocol. Finally to prevent the Black hole attack, a Counter Algorithm is proposed to make the AODV Protocol secure. Performance of the protocol was measured using metrics like Packet Delivery Ratio and Average End-to-End Delay. A method to identify the malicious node was explained in order to avoid forwarding of the information to the malicious node in the routing table. The solution given did not impose any overhead on the nodes in the network.

1.1.3. Temporally Ordered Routing Algorithm

The Temporally ordered Routing Algorithm is outlined considering the connection inversion idea and is a calculation free of circles [12]. This convention works in three stages as takes after: (a) The principal stage begins with making a route (b) the second stage runs with keeping up the route and (c) the third stage evacuates all the invalid routes .Comparison of TORA protocol with other routing protocols like OLSR, DSDV with the metrics like Throughput, Control Overhead, End to End Delay and Packet Delivery Ratio and was seen that it is better in performance with varying number of nodes using NS-2 simulator.

1.1.4. Associativity Based Routing

The Associativity Based Routing protocol does not have loops, free of deadlock and no duplicate packets [13]. It focuses on route longevity. As there are very few broken communication links and less need for reconstruction of the routes the overhead involved is less. An improved version of ABR was to optimize the bandwidth and demand to reduce the overhead based on the position information was proposed. It was concluded that the path setup time was long for the routes which gave a scope for the future research to improve the ABR Protocol.

1.1.5. Signal Stability-based Adaptive Routing Protocol

The working of SSR routing protocol states that the large routing tables are not required for routing [14]. The network will not be congested with the control messages but a type of denial of service attack is a threat to this protocol. The Signal Stability Table is maintained that has information of signal's strength of all nearby nodes. The protocol was simulated in OmNet and a metric known as CPU usage was considered to measure the performance. It proved that when there are malicious nodes the usage of CPU was more than in the absence of malicious nodes.

*2.2 Reasons for Threats and Attacks*

Reasons for threats have been summarized as shown in the following figure.

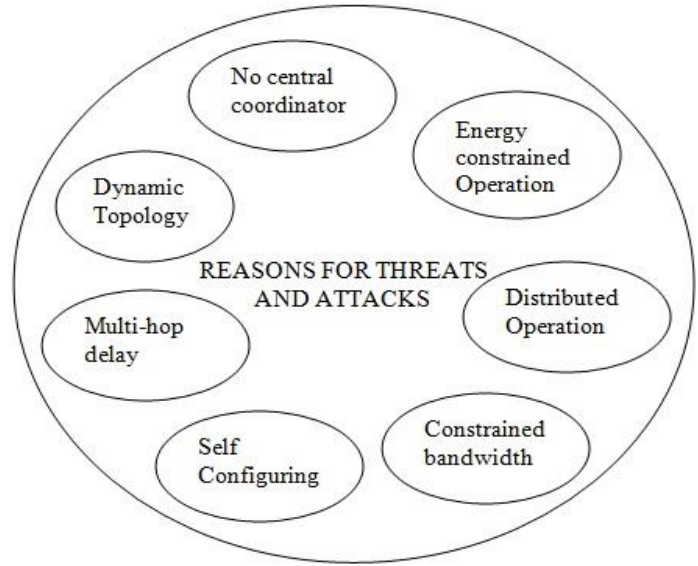

**Figure 4.** Reasons for Threats and attacks.

**3. Methodology**

The main aim of this research is to identify the various characteristics and approaches of the reactive routing protocols and study the attacks and the reasons for such attacks in reactive routing protocols in order to propose a general framework to avoid such attacks or withstand effectively from the same and propose an INESH algorithm to identify the ideal node to be used in forwarding the packets. The study aims to consider the reactive routing protocols with different types of attacks on them along with

- Studying the reasons for threats and attacks along with malicious behaviour of the nodes in the network.
- The effect of forwarding packets bearing threats and attacks in the network and eliminating them from the network
- After the reasons for threats and attacks have been identified, a generalized framework to be proposed to enhance the reactive routing protocols with high security and better performance.

The study mainly uses the following reactive routing protocols
- DSR (Dynamic Source Routing Protocol)
- AODV (Ad hoc On Demand Distance Vector Routing Protocol)

An analytical Research Methodology has been adopted in conducting the study. The various routing protocols have their own methodologies to send the information to the destination. This research considers the reasons for the threats and attacks in the reactive routing protocols and the effect of eliminating the nodes from the path due to the malicious behavior of nodes. The plan is to propose a strategy which will help in maximizing the throughput and minimizing the routing overhead on the routing protocols and thus help in the selection of an ideal node on the most optimal routing path for any protocol to send the information to the destination.

**4. Discussions**

The routing protocols DSR and AODV have been simulated using a simulator namely ASIM (Ad hoc Simulator) with the varying number of nodes to know the effects when there are malicious nodes in the network using the following constraints.

**Table 1.** Simulation Parameters

| Parameter | Values |
|---|---|
| Number of Nodes | 35, 40, 45, 50 |
| Physical Terrain Dimensions | 500m x 550m |
| Communication Range | 150 m |
| Packet Size | 120 bits |
| Maximum Speed | 20m/s |
| Source Data Rate | 4 packets/sec |
| Application Data Packet Size | 512bytes |

The following graphs have been depicted with the time in seconds on x-axis and throughput on y-axis for DSR and AODV protocols.

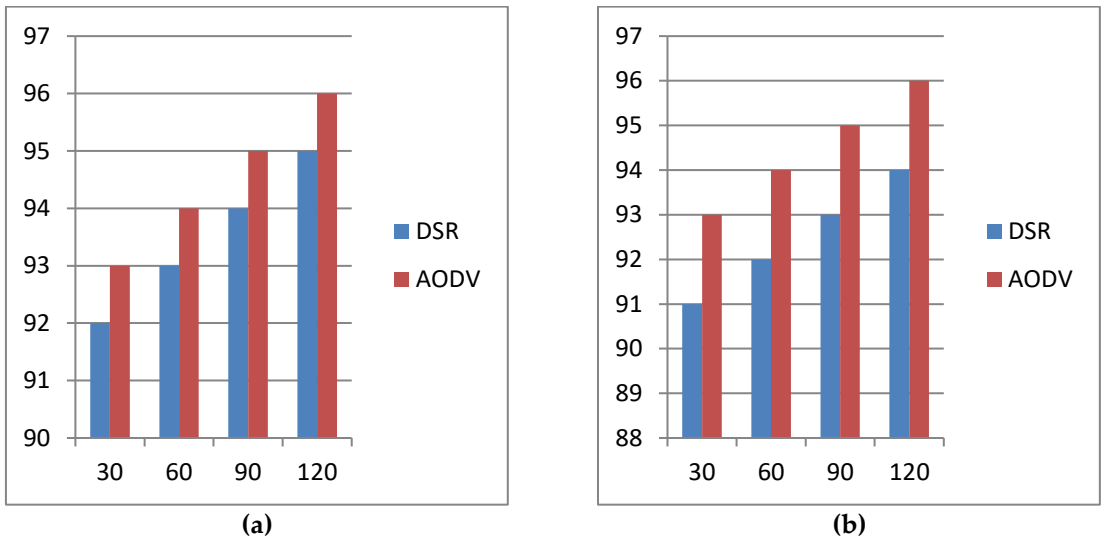

**Figure 5.** Number of Nodes as 35 on the left and 40 on the right.

In figure 5, the throughput is less when the time is less but it increases as the time given to each protocol increases. The reason is when more time is given the protocols find the alternate path where there can be malicious nodes in its path to the destination resulting in more throughput.

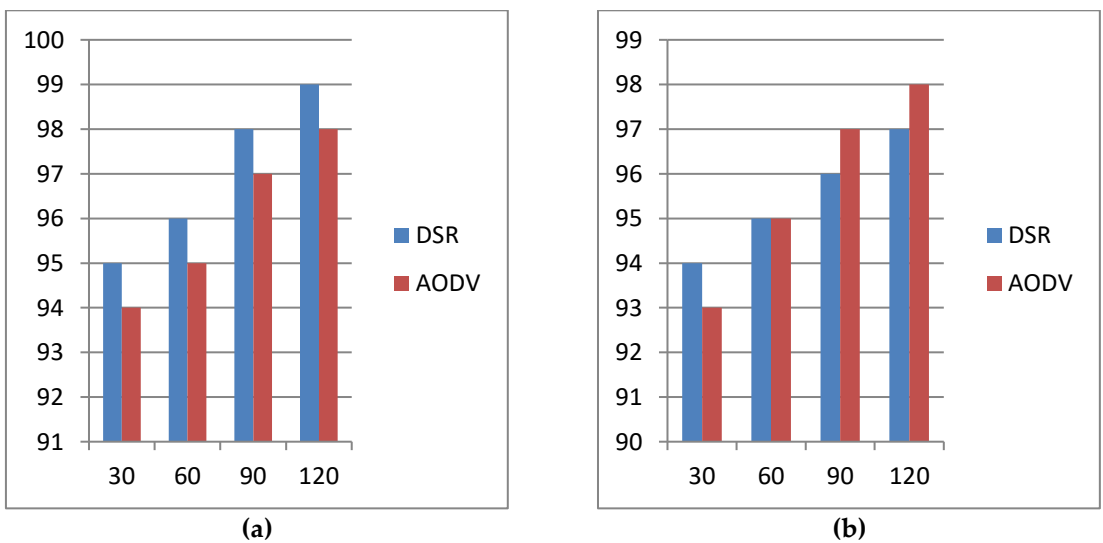

**Figure 6.** Number of Nodes as 45 on the left and 50 on the right.

In figure 6, the throughput in both the protocols is still more increased with the same reason that more number of alternate routes is available to reach the destination.

From both the above figures, the reason for having high throughput was studied and the following issues have been identified.

i. Nodes as hosts and routers: Because the nodes are assuming a double part of hosts and furthermore as a router in sending the parcels, if some node turns into a noxious node, there has been a double drawback by abusing the movement and by dropping the messages.

ii. Scarcity of Resources: To make the data secured while transmission by utilization of cryptographic calculations makes it confused to be actualized in an infrastructure less system as the resources are exceptionally constrained as opposed to the framework organize.

iii. Mobility: The dynamic changing topology of system offers degree to the flawed nodes more shots for attacks.

If the malicious nodes will be eliminated from the path then the throughput will still more decrease as the nodes tend to waste time in finding the next neighbor node on its path to reach destination. In some cases though there is a trusted node, it may give feel to the sender that it is a malicious node and might not forward the packet to it. And sometimes when continuously malicious nodes are deleted then worst case is that there might not be a path to send the packet to destination. Therefore a route request again has to be initiated by the routing protocol. Instead, the following algorithm can be run whenever a node has to find an intermediate node.

*4.1 Procedure INESH*

The following algorithm can be run at any node whenever a route request is initiated, thereby reducing the load on the protocol itself to find the trust worthy node as well as the shortest path to send the packets to the destination. So the protocol can perform its intended operation without focusing on finding the correct neighbor.

(N: set of nodes 1... n {Source as Node 1, Destination as Node n}
Create an adjacency lists;
Establish cost between two nodes(u, w): using cost functions;)
Set distance between pair of nodes;
Trust(i, j): values from 0 to 1{0 as minimum, 1 as maximum}
Begin:
Initialize all the variables;
   Create an empty Priority Queue;
sDist[1]←0;         {The distance to the source is zero}
for all Nodes Ni in N – {1} do      {no edges have been explored yet}
sDist[w]←∞ end for;
Fill the queue with Nodes w in N organized by priorities sDist[w]; endInitialize;
Repeat for all nodes
If trust((Ni>Nj) && (First hand info from the neighbor node is negative))← Delete Node Ni from the path;        {Increment the trust value of Nj and update the routing table}
For all of the neighbors w in Adj[v] compute the shortest path using any algorithm
update w in queue
endif
endfor until Queue is empty
end INESH;.

**4. Conclusion and Future Enhancements**

The proposed integrated approach at every node reduces the load on protocol itself and helps to improves the throughput with low network overhead though there are malicious nodes in the dynamic network. The INESH algorithm inclusion not only allows to find the correct neighbor but also helps in identifying the malicious node through the path. As the nodes are not eliminated from the network unless and until they fail in proving their trust, the network is strong and will result in more throughput with low network overload. In future changes in these protocols can be done by using this approach and a simulative research can be done.

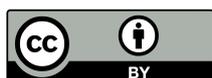